\documentstyle[preprint,aps]{revtex}
\input epsf
\begin{document}
\draft
\title{Parallel Invaded Cluster Algorithm for the Ising Model}  
\author{Y. S. Choi and J.
Machta}
\address{Department of Physics and Astronomy, University of
Massachusetts,  Amherst, MA 01003-3720}
\author{P. Tamayo}
\address{Thinking Machines Corp., Bedford, MA 01730} 
\author{L. X. Chayes}
\address{Department of Mathematics, University of California, Los
Angeles, CA 90095-1555}
\narrowtext 
\tightenlines
\date{\today}
\maketitle
\begin{abstract}
A parallel version of the invaded cluster algorithm is described.
Results from large scale (up to $4096^2$ and $512^3$) simulations of the
Ising model are reported.  No evidence of critical slowing down is found
for the three-dimensional Ising model.  The magnetic exponent is estimated
to be $2.482 \pm .001$($\beta/\nu=0.518\pm .001$) for the three-dimensional
Ising model.

 \end{abstract}

\section{Introduction}
 
Advances in algorithms and hardware have greatly increased the
power and scope of Monte Carlo studies of critical phenomena. 
Cluster algorithms, first introduced by Swendsen and Wang \cite{SwWa},
largely overcome the problems of critical slowing and thereby permit the
study of very large systems using a reasonable number of Monte Carlo steps. 
Parallel machines provide platforms on which very large systems may
be efficiently studied.  Several groups
\cite{samp,FlTa,GuTa} have implemented
cluster algorithms on parallel machines to obtain high precision results
for the three-dimensional Ising model.

The invaded cluster (IC) algorithm \cite{MaCh95,MaCh96} is a new type of
cluster algorithm with features that make it very attractive
for high precision studies of critical phenomena.  First, it has less
critical slowing than other cluster algorithms.  Indeed, here we provide
strong evidence that for the three-dimensional Ising model there is no
critical slowing down.  Second, the IC algorithm does not use an estimate
of the critical temperature as an input;  instead, the critical point is
sampled without any fine-tuning of parameters and an estimate of the
critical temperature is obtained as an {\em output} of the simulation.  On
the other hand, the IC algorithm samples the IC ensemble rather than the
canonical ensemble.  Although the two ensembles are believed to be
equivalent in the infinite volume limit, the finite-size scaling
properties of the IC ensemble are not yet well-understood.  Here we
present results that test a simple finite-size scaling hypothesis and
determine the associated finite-size scaling exponent.

In this paper we describe an implementation of the invaded cluster
algorithm for distributed memory SPMD (single program
multiple data) parallel machines.  Using this algorithm we
studied two- and three-dimensional Ising systems up to size
4096$^2$ and 512$^3$, respectively.  There are several motivation for this
project.  The first is to develop an efficient implementation of the IC
method for parallel computation.  This is a non-trivial
problem because the IC method appears to be quite
sequential.  The second objective is to better
understand the dynamics of the IC algorithm and the finite-size scaling
properties of the IC ensemble.  These properties are
interesting in their own right and they must be understood before the
algorithm can be used for high-precision studies.  Finally, we wish to
obtain accurate results for the critical temperature and exponents
for the three-dimensional Ising model.  With relatively modest effort we
have been able to obtain accuracy comparable to recent studies in the
measurement of $\beta/\nu$  and somewhat less accuracy for the critical
temperature. We argue that the IC method is likely to be the most
efficient method for obtaining very high precision estimates for the
critical temperature and the magnetic exponent for the Ising model and
related spin systems.

In Sec.\ \ref{sec:ic} we describe the invaded cluster method and discuss
our parallel implementation.   In Sec.\ \ref{sec:per} we discuss the 
performance of the parallel IC algorithm running on CM-5 
machines. In Sec.\ \ref{sec:num} we present some high-precision  numerical
results for the three-dimensional Ising model using the parallel IC
algorithm. Conclusions are presented in Sec.\ \ref{sec:con}. 

\section{Invaded Cluster Algorithm}
\label{sec:ic}
\subsection {Cluster Methods}
In cluster Monte Carlo methods a spin system is sampled in both its
spin representation and in an associated {\it graphical
representation}. The graphical representation is an interacting
percolation model that is usually defined on the bonds of the
lattice.  The graphical representation by itself contains all of the
information relevant for the study of the spin system.  A single
Monte Carlo (MC) step consists of two parts, a `bond move' and a
`spin move.'  During the bond move, the spins are frozen and the
graphical configuration (bond configuration) is updated in accord
with the current spin configuration.  During the spin move, it is
the opposite:  the bonds are frozen and the spin configuration is
updated. In a critical region, tremendous improvements in the
efficiency of cluster methods over traditional local dynamics can
arise if the graphical representation has a coinciding percolation
threshold.  In this case, clusters of spins on all scales are
coherently updated.

Since the IC and Swendsen-Wang (SW) algorithms are very similar let
us begin with a brief description of the SW algorithm. Consider the
nearest neighbor Ising magnet on the square or cubic lattice defined
by the Hamiltonian
\begin{equation}
\beta H = -K\sum_{\langle i,j \rangle}\sigma_i\sigma_j
\end{equation}
where $\sigma_i = \pm 1$, $K$ is the dimensionless coupling, and the sum
is over all the bonds of the lattice. During the bond move, each
`satisfied' bond ($\langle i,j \rangle$ is satisfied if
$\sigma_i=\sigma_j$) is independently occupied with probability $p$ or left
vacant with probability $(1-p)$. Unsatisfied bonds are never occupied. 
After a bond configuration has been created in this fashion, {\it
clusters} are identified: A cluster is a set of (like) spins connected by
occupied bonds; an isolated spin is also counted as clusters.  Next, the
spin configuration is erased and the bond move is completed. The spin
move for the SW algorithm consists of assigning either spin--up or
spin--down with probability 1/2 independently to every cluster.  Each
spin in the cluster takes the spin value assigned to the cluster and
then the bond configuration is erased and the spin move is completed.
The SW algorithm samples the canonical ensemble at coupling $K$
where $K$ is related to $p$ via
\begin{equation}
p=p(K) \equiv 1 - e^{-2K}.
\end{equation}

The spin move for the IC algorithm is the same as for the SW
algorithm:  each cluster is assigned either spin--up or spin--down
with probability 1/2.  The IC bond move consists of occupying the
satisfied bonds in random order until the cluster configuration
first fulfills a stopping rule.  During the bond move,
bonds are occupied one at a time and after each bond is added, the
cluster structure is checked to see if the stopping rule is
satisfied.  There are a number of useful stopping rules, some of
which, evidently, cause the algorithm to sample the Ising critical
point. In this study, we will use `topological' rules which are
appropriate for systems with periodic boundary conditions.  There
are several versions: The `1-span' topological rule demands that
some cluster is connected around the system in at least one of the
periodic directions -- as soon as this condition is fulfilled, the
bond move is complete and no more bonds are added to the
configuration.   In general, the $k$-span rule ($k \leq d$) requires
that there is connectivity around the system in at least $k$
directions. Although all of the $k$-span rules sample the Ising
critical point, the different rules have different finite-size
scaling behavior.

The critical coupling and the magnetic exponent may be obtained
in the IC approach.  We begin with $K_c$. Let $f$ be the ratio of
the number bonds occupied during the bond move to the number of
satisfied bonds for a single Monte Carlo (MC) step. The average value
of $f$ is an estimator of the critical temperature according to the
relation
\begin{equation}
\label{eq:fbeta}
\lim_{L \rightarrow \infty} <\!f\!> =  1 - e^{-2K_c}
\end{equation}
where $K_c$ is the critical coupling and $L$ is the system size.
To understand this relation, note that the only difference between
the SW and IC algorithms is the way in which $f$ is determined in a
bond move.  For the SW algorithm, each satisfied bond is
independently occupied with probability $p$ so that the ratio of
occupied to satisfied bonds is a binomial random variable.  For the
IC algorithm, the stopping rule determines the number of
occupied bonds.  However, for large systems, one can presume that the
fluctuations in $f$ become small for both algorithms. As a
consequence, the IC algorithm that outputs $<\!f\!>$  yields the
same value for all local observables as the SW algorithm with
$p(K)=<\!f\!>$ so that the relation $<\!f\!> \simeq 1 - e^{-2K}$
yields an estimate of the temperature of the IC ensemble. The bond
percolation threshold on the set of satisfied bonds at the Ising
critical point is known to be $p(K_c)$.  Since the $k$-span rules
forces the cluster configuration to the percolation threshold, we
see that Eq.\ (\ref{eq:fbeta})
holds.  This argument (which still lacks a rigorous proof) is
discussed in more detail in Ref.\ \cite{MaCh96}

Let us now turn to a discussion of the magnetic exponent.  For a
critical stopping rule, such as a topological rule, the cluster that
fulfills the stopping condition is called the {\it spanning
cluster}.  This spanning cluster can be thought of as a typical
example of a large--scale cluster at criticality and thus its fractal
dimension yields the magnetic exponent for the system. In
particular, if $M$ is the number of sites in the spanning cluster,
we determine the fractal dimension $D$ via
\begin{equation}
\label{eq:D}
M \sim L^D 
\end{equation}
The fractal dimension is related to other critical exponents via $D
= y_h = d- \beta/\nu$ where $y_h$ is the magnetic exponent, $\beta$
the magnetization  exponent, $\nu$ the correlation length exponent
and $d$ the spatial dimension.  At the present time we do not know
how to extract the thermal exponent from the IC algorithm.

The finite-size scaling properties of the IC ensemble are currently
under investigation.  Here we state the fundamental assumption that
will be used later in the analysis of the data.   Let ${\cal
F}(f,L)$ be the (cumulative) distribution function for $f$ measured
in a system of size $L$.  We suppose that there is an exponent, $u$
and function ${\cal G}$ such that
\begin{equation}
\label{eq:fss}
{\cal F}(f,L) \sim {\cal G}[(f-p(K_c))L^u]
\end{equation}
Naive finite-size scaling arguments suggest that $u=1/\nu$ with
$\nu$ the Ising correlation length exponent, however it is quite
apparent that this is not the case.  Measurements of $u$ reported in
Ref.\ \cite{MaCh96} and below in Sec.\ \ref{sec:num} yield a value
$u \approx 0.69$ in $d = 3$.  Presumably this $u$ is the inverse of the
correlation length exponent for a related percolation--Ising hybrid model
but at present, no definitive conclusions have been obtained.

\subsection {Parallelizing the Invaded Cluster Algorithm}

The approach taken here to parallelize the IC algorithm is based
upon the parallel SW algorithm described in Ref.\ \cite{FlTa}.  The basic
idea for both algorithms is to divide the lattice into cells each of which
is handled by a single processor. During the bond move the cluster
configuration must be determined.  Every site of the lattice has an
integer label that identifies its cluster membership.  Two sites are in
the same cluster if they have the same label. At the beginning of the bond
move, every site has a unique local label.  The SW bond move proceeds in
two stages.  In the {\em local} stage, each processor in parallel occupies
a fraction $p$ of the satisfied bonds in its cell and identifies the local
cluster structure using a conventional sequential method. The local
clusters are the clusters that would exist if no bonds on the boundaries
of the cell were occupied.  During the  {\em relaxation} stage, global
labels are created and global clusters are identified.    
Global clusters which are formed out of local clusters by boundary
connections between cells are assigned the lowest valued label from the
the labels of the connected local clusters.  This is accomplished
iteratively by the exchange of labels between neighboring cells.
A very similar relaxation stage is used for the IC algorithm except that
more information must be exchanged between nodes to allow for checking the
stopping condition.  

Since communication between processing nodes is much slower than
processes handled by a single node it is desirable that
individual nodes do as much work as possible between
communication steps. On the other hand, the sequential IC algorithm checks
for spanning after each new bond is occupied.  Clearly this would not
be a good approach to directly parallelize.  Instead, the algorithm
converges to the value of $f$ by a sequence of choices of $p$. 
Imagine that an independent uniform random number between 0 and 1 is
assigned to each bond of the lattice.  (In fact, this is not quite the
approach that is actually used but, for purposes of explanation, suppose
that this is the case.)  On the first step each processor provisionally
occupies all satisfied bonds with random numbers less than $p_1$.  The
resulting cluster configuration is identified and the stopping rule is
checked.  For the sake of argument, suppose that spanning is not
detected.  In this case, the provisionally occupied bonds are permanently 
occupied and a new value $p_2>p_1$ is chosen.  The current cluster
configuration is stored and additional bonds are provisionally occupied
with random numbers from $p_1$ to $p_2$. The cluster configuration is
modified accordingly and the stopping condition is checked.  If spanning
is not detected, the new cluster configuration is stored and the
provisionally occupied bonds given tenure.   The value of $p$ is
incremented in this way until on the $s$th step the stopping condition is
fulfilled.  At this point we know that spanning occurs between $p_{s-1}$
and $p_s$.  We backtrack to the cluster configuration (saved in memory)
associated with $p_{s-1}$ and try again, this time choosing $p=(p_{s-1} +
p_s)/2$.  Henceforth, the algorithm carries out a binary search, each time
reducing the range of $p$ by half until the stopping condition is just
fulfilled and $f$ is determined.

In the above discussion we supposed that random numbers
are assigned to each bond and that on each step all
satisfied bonds with random numbers less than $p$ are occupied. 
This is not an efficient procedure because it requires examining
every bond each time $p$ is changed.  Instead we do the
following.  Each processor creates a random permutation of the $n$
bonds in its cell. This permutation determines the
order in which bonds will be tested by the processor.  On the first
step, processor $i$ tests its first $h_i^{(1)}$ bonds where $h_i^{(1)}$ is chosen
from the binomial distribution, ${\cal B}[n,p_1]$ for $n$ trials with
success probability $p_1$.  This is statistically equivalent to testing
all bonds with random numbers less than $p_1$ as described in the
previous paragraph. Suppose that spanning is not found, then $p_1$ is
incremented to $p_2$.  At this point processor $i$ has permanently
occupied $a_i^{(1)}\leftarrow h_i^{(1)}$ bonds.  Processor $i$ now tests
the next $h_i^{(2)}$ bonds where now $h_i^{(2)}$ is chosen from ${\cal
B}[n-a_i^{(1)},(p_2-p_1)/(1-p_1)]$. This process of
occupying bonds continues as long as the stopping condition is not
fulfilled and on the $m$th such step the number of occupied bonds is
incremented according to $a_i^{(m)} \leftarrow a_i^{(m-1)} +
h_i^{(m)}$.    Suppose that spanning is first detected on step $s$ and
that $h_i^{(s)}$ bonds have been provisionally occupied by processor $i$
during this step.  On the next step each processor  tries to occupy a
smaller number of bonds corresponding to splitting the difference between
$p_{s-1}$ and $p_s$. To do this, processor $i$ tests the next
$h_i^{(s+1)}$ bonds starting with the $(a_i^{(s-1)}+1)$th bond where 
$h_i^{(s+1)}$ is chosen from ${\cal B}[h_i^{(s)},1/2]$.  A little
reflection shows that this procedure is statistically equivalent to the
one described above for carrying out the binary search.  The advantage of
this approach is that on successive steps it is not necessary to check
every bond in the cell. Note that the local ordering of bonds is fully
specified at the outset of the bond move but that the global ordering of
the bonds is generated as needed and never fully specified.

\section{Performance}\label{sec:per}
 
\begin{figure}[h]
\epsfxsize = 3.5in
\begin{center}
\leavevmode
\epsffile{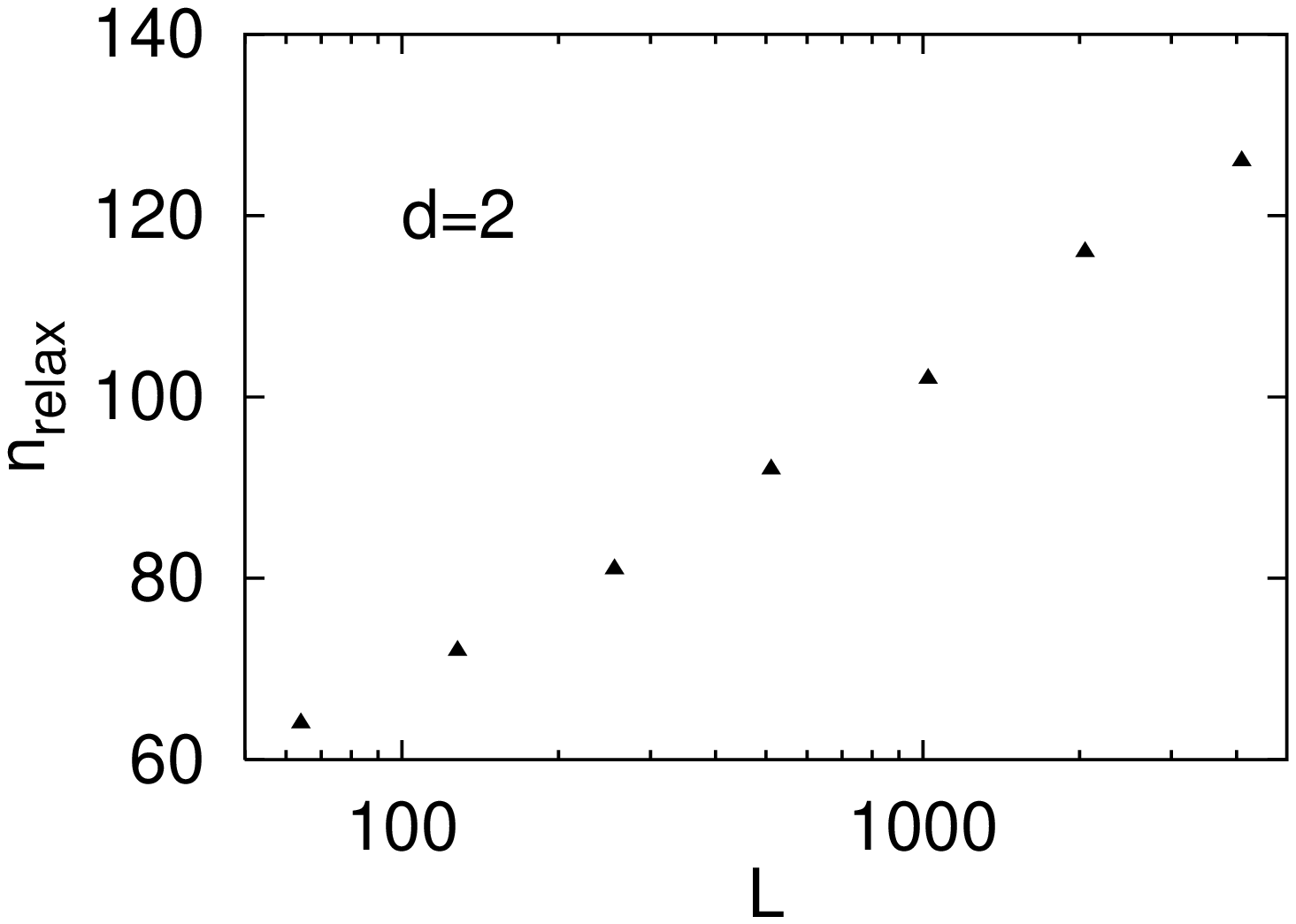}
\caption{The number of relaxation cycles per MC step
$n_{\rm relax}$ vs.\ system size $L$ for the two-dimensional Ising
model.} \label{fig:fig1}
\end{center}
\end{figure}

\begin{figure}[h]
\epsfxsize = 3.5in
\begin{center}
\leavevmode
\epsffile{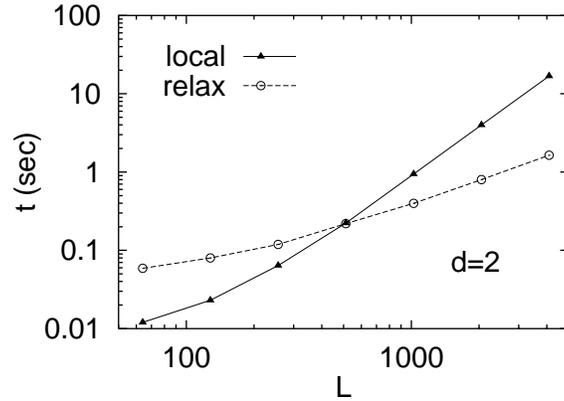}
\caption{Local time and relaxation time per MC step vs.\
system size  $L$ for the two-dimensional Ising model.}
\label{fig:fig2}
\end{center}
\end{figure}

We tested the parallel IC algorithm for the two-dimensional Ising model 
on a 32-node CM-5 machine. Fig.\ \ref{fig:fig1} shows that the number of 
relaxation cycles $n_{\rm relax}$ per MC step as a function of system
size $L$. The roughly logarithmic increase in  $n_{\rm relax}$ results from
the binary search method used to find $f$. We separately measured the time
spent in local cluster growth $t_{\rm local}$ and relaxation
$t_{\rm relax}$. The local time increases as the number of spins per node,
thus, for a fixed number of processors, $t_{\rm local} \sim  L^d$.  
The relaxation time is dominated by
communication between nodes so $t_{\rm relax}$ is proportional to the
boundary area of the local lattice and the number of relaxation cycles,
$t_{\rm relax} \sim L^{d-1}\log L$.  Fig.\ \ref{fig:fig2} shows local and
relaxation times as a function of system size.  The relaxation time is
greater for small systems  but, when the number of spins per node exceeds
several hundred thousand the local time dominates. 

\begin{figure}[h]
\epsfxsize = 3.5in
\begin{center}
\leavevmode
\epsffile{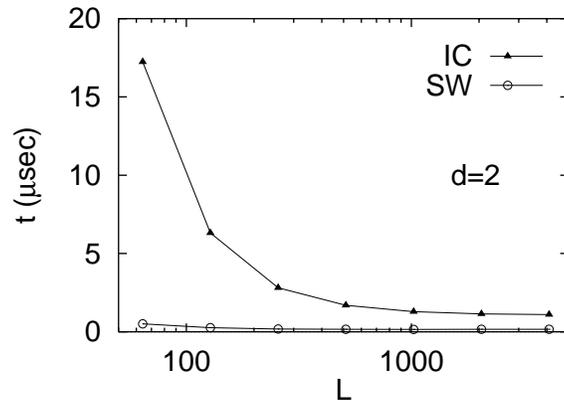}
\caption{Time per spin per MC step for the two-dimensional Ising
model vs.\ $L$ for the parallel
IC and SW algorithms.}
\label{fig:fig3}
\end{center}
\end{figure}

In Fig.\ \ref{fig:fig3} we compare the performance of the parallel IC
algorithm with that of the parallel SW algorithm for the two-dimensional
Ising model.  The total time per spin per MC step for both algorithms
decreases as the system size increases and saturates for large systems. 
For large systems in both two and three dimensions, the
time for one MC step for the parallel IC algorithm is about a factor of $7$
larger than for the parallel SW algorithm. 

We obtained  $1.1\ \mu \:$s per spin per MC step for the parallel IC
algorithm for the two-dimensional Ising models with  system size $4096^2$
on the 32-node CM-5 machine and $3.4\ \mu  \:$s per spin per MC step for
the three-dimensional Ising model on the same machine and with the same
number of spins ($256^3$). The slower speed for three-dimensions results
from the need for more communications between nodes since each node
has six neighbors rather than four. For both two and three dimensions the
time per spin per MC step decreases as the system size increases. The
relaxation time is, however, still comparable to the local time for
$256^3$, so the parallel IC algorithm for the three-dimensional Ising
model can be expected to become more efficient as the system size
increases holding the number of processors fixed.

To test the sensitivity of the algorithm to the random number
generator, we performed simulations using two different random number
generators.  Most of the data is collected using the 250-shift
generator in which new random numbers are produced by
$X_i=X_{i-103}\oplus X_{i-250}$, where $X_i$ are 32-bit
integers  and $\oplus$ represents the bitwise exclusive OR
operation.  We also tested a
combined shift register generator, introduced in Ref.\ \cite{BlLuHe}. The 
combined generator consists of two shift registers  $a_i=a_{i-9218}\oplus
a_{i-9689}$ and $b_i=b_{i-97}\oplus  b_{i-127}$ with 32-bit integers $a_i,
b_i$. New random numbers of the  generator are combined by $X_i=a_i\oplus
b_i$. We observed no difference between results from the
two generators. For example, the mean value  of $f$ for system size
$80^3$ was measured as $.358\ 144 \pm  .000\ 006$ for
the first generator and $.358\ 143\pm .000\ 003$ for the second.

\section{Numerical Results}
\label{sec:num}

We studied the three-dimensional Ising model using  the parallel IC
algorithm for system sizes $16^3$ to $512^3$. Unless otherwise stated we
used the 1-span rule.   For each run we allowed 200 MC steps for equilibration
before collecting data.  Independent runs consisted of $10\ 000$ MC steps
for the smallest systems (up to $96^3$)  to  $500$ MC steps for the
largest system (512$^3$).  For each size, a number of independent runs
were made.  We obtained statistics for $f$ and the size of the spanning
cluster.  Error bars represent the standard error (one standard deviation)
associated with these independent runs.  The data for the 1-span rule is
given in Table I.

\begin{figure}[h]
\epsfxsize = 3.5in
\begin{center}
\leavevmode
\epsffile{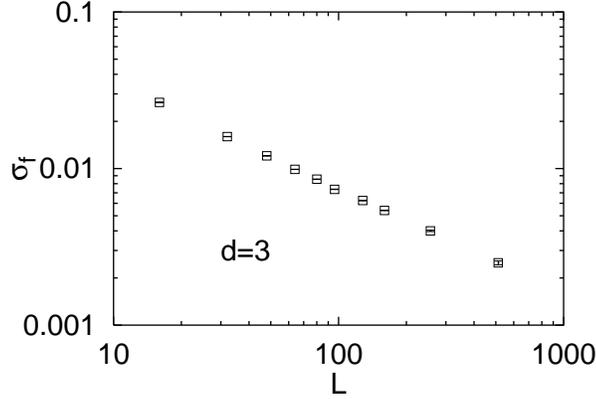}
\caption{Double logarithmic plot of $\sigma_f$ vs.\ system size $L$
for the three-dimensional Ising model. The slope of the curve is $.69\pm
.01$ from a linear fit.}
\label{fig:fig4}
\end{center}
\end{figure}

Figure \ref{fig:fig4} is a double logarithmic plot of
$\sigma_f=(<\!f^2\!>-<\!f\!>^2)^{1/2}$ vs.\ system size. A fit of the form
$\sigma_f=a + bL^{-u}$, yields $a=0.0003 \pm .0002$ and $u=0.69 \pm .01$.
This result strongly supports the hypothesis that  $\sigma_f$ approaches
zero as a power law.     

\begin{figure}[h]
\epsfxsize = 3.5in
\begin{center}
\leavevmode
\epsffile{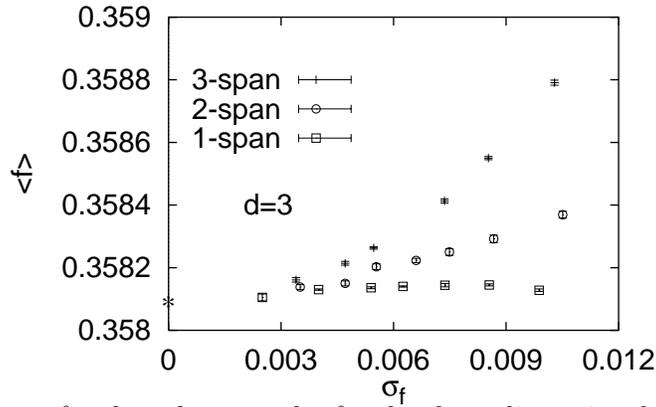}
\caption{Plot of $f$ vs.\ $\sigma_f$ for three $k$-span rules for the
three-dimensional Ising model.  The point on the vertical axis is
the accepted infinite volume estimate.} 
\label{fig:fig8}
\end{center}
\end{figure}

The finite-size scaling hypothesis, Eq.\ (\ref{eq:fss}) implies an
asymptotically linear relation between $<\!f\!>$ and $\sigma_f$.  Figure
\ref{fig:fig8} shows plots of $<\!f\!>$ against $\sigma_f$ for the
three $k$-span rules.  Although these plots are
consistent with asymptotic linear behavior, it is clear that
there are large subleading corrections to scaling.  For the
1-span rule these corrections are negative and for the 3-span rule
they are positive.  

\begin{figure}[h]
\epsfxsize = 3.5in
\begin{center}
\leavevmode
\epsffile{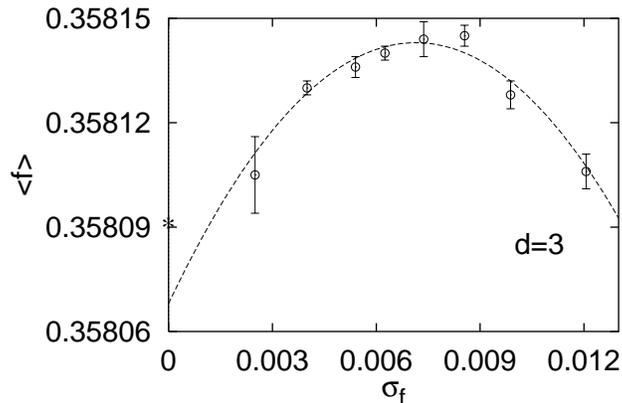}
\caption{$<\!f\!>$ vs.\  $\sigma_f$ for the 
three-dimensional Ising model using the 1-span rule. The dotted
line is a quadratic fit and yields $p(K_c)=0.358\ 068 \pm
.000009$.  The point on the vertical axis is
the accepted infinite volume estimate.}   
\label{fig:fig5} \end{center} \end{figure}

Figure \ref{fig:fig5} shows more detail for the 1-span rule where,
over the range of system sizes studied, $<\!f\!>$ varies only in
the fifth significant digit.  However, to obtain yet higher accuracy
than $p(K_c)=.3581$ requires either larger system
sizes or a good fit to the subleading terms.  Fitting to the form
$<\!f\!>=p(K_c) + a \sigma_f + b \sigma_f^2$ we obtain $p(K_c)=0.358\
068\pm .000\ 009$ with $a=.021$ and $b=-1.47$. The corresponding
critical coupling, quoted with two standard deviation errors, is
$K_c=0.221\ 637 \pm .000016$.   The fit is shown as a dotted line in the
figure.  The primary difficulty in obtaining a high precision value of the
critical coupling is the relatively large corrections to scaling,
represented by the fact that the fitting parameter $b$ is a good deal
larger than $a$.  Our estimate of the critical coupling is consistent with
recent values~\cite{GuTa,FeLa,BlLuHe} (e.g. $K_c=0.221\ 654\ 6 \pm .000\
001$ \cite{BlLuHe}) but the error is much larger.   

\begin{figure}[h]
\epsfxsize = 3.5in
\begin{center}
\leavevmode
\epsffile{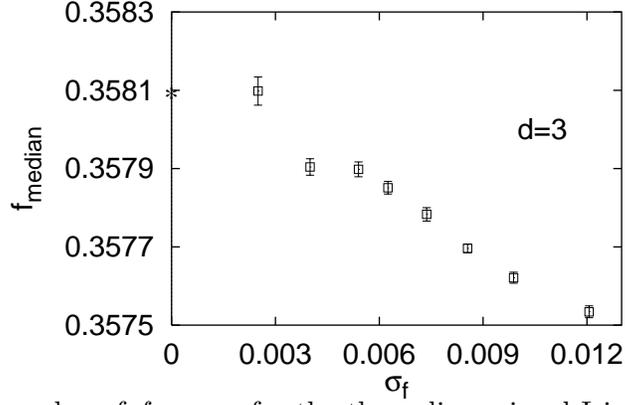}
\caption{The median value of $f$ vs.\  $\sigma_f$ for the 
three-dimensional Ising model using the 1-span rule.}  
\label{fig:median}
\end{center}
\end{figure}

Figure \ref{fig:median} shows the median value of $f$ for the 1-span rule
versus $\sigma_f$.  According to the finite-size scaling hypothesis this
plot should also be linear and, indeed, approximately linear behavior is
evidenced over the whole range of system sizes however it should be noted
that the variation in the median is much larger than for the mean. 
Furthermore, errors in the mean are less.  Thus appears that the mean is
a better estimator of the critical coupling. 

\begin{figure}[h]
\epsfxsize = 3.5in
\begin{center}
\leavevmode
\epsffile{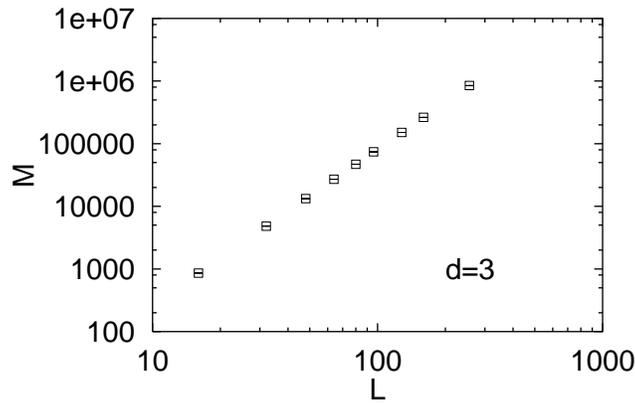}
\caption{Double logarithmic plot of the mass of the spanning cluster,
$M$ vs.\ system size $L$ for the three-dimensional Ising model.}
\label{fig:fig7}
\end{center}
\end{figure}

We measured the mass (the number of sites) $M$ of the
spanning  clusters. Using Eq.\ (\ref{eq:D}) we can find the fractal
dimension of the spanning cluster and the magnetic exponent. The double
logarithmic plot  of $M$ and $L$ in Fig.\ \ref{fig:fig7} shows that
the data is well described by a power law and an examination of
different ranges of system size yields no clear trends in the power.  A
linear fit for $L \geq 48$ yields
$\beta / \nu = 0.518\pm .001$ consistent with a recent value $0.5185\pm
.0008$ \cite{BlLuHe}.

\begin{figure}[h]
\epsfxsize = 3.5in
\begin{center}
\leavevmode
\epsffile{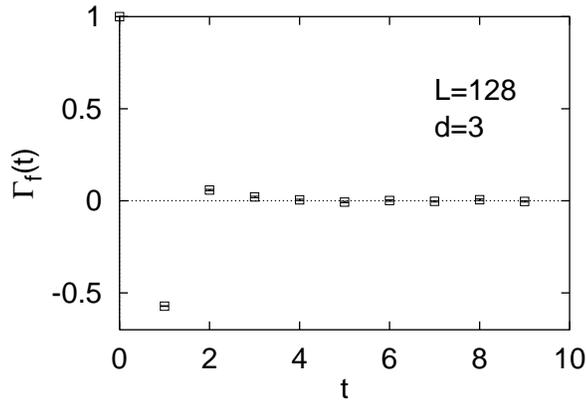}
\caption{
Autocorrelation function for $f$, $\Gamma_f$ vs.\ $t$ for the
three-dimensional Ising model for system size $128^3$.}
\label{fig:fig9}
\end{center}
\end{figure}

Next we consider the dynamical properties of the algorithm. The
normalized autocorrelation function of an observable $A$ is defined
by, 
\begin{equation}
 \Gamma_A(t)=\frac{<\!A_0 A_{t}\!>-<\!A\!>^2}{<\!A^2\!>-<\!A\!>^2},
\end{equation} 
where $t$ is time in Monte Carlo steps. Figure
\ref{fig:fig9} shows the autocorrelation function for $f$ for a
128$^3$ system.  The striking feature of this curve is that it is
negative on time step 1 indicating anti-correlations between
successive MC steps.  As discussed in Refs.\ \cite{MaCh95,MaCh96,LiGl}
these anti-correlations are associated with the negative feedback mechanism
that forces IC dynamics to the critical point. 
By the fourth step it is not possible to distinguish the autocorrelation
function from zero. 

The integrated autocorrelation time $\tau_A$ is
defined by, 
\begin{mathletters} 
\begin{equation}
 \tau_A(w)=\frac{1}{2}+\sum_{t=1}^w\Gamma_A(t), 
\end{equation}
\begin{equation}
 \tau_A=\lim_{w\rightarrow\infty}\tau_A(w). 
\end{equation}
\end{mathletters}

\begin{figure}[h]
\epsfxsize = 3.5in
\begin{center}
\leavevmode
\epsffile{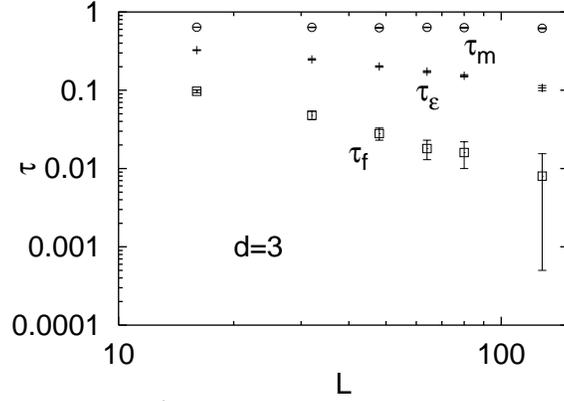}
\caption{Double logarithmic plot of autocorrelation times $\tau_m,
\tau_{\epsilon}, \tau_f$ vs.\ system size $L$ for the
three-dimensional Ising model.}
\label{fig:fig6}
\end{center}
\end{figure}

Using the procedure described in Ref.\ \cite{MaCh96}, we measured 
integrated autocorrelation times for the magnetization $m$,
energy $\epsilon$ and temperature estimator $f$.  The autocorrelation times
are given in Table II and plotted against system size
in Fig.\ \ref{fig:fig6}.  Due to these
anti-correlations  $\tau_f$ and $\tau_{\epsilon}$ {\em decrease} with
system size; $\tau_m$ remains nearly constant.  Thus IC dynamics for the
three-dimensional Ising model apparently does not suffer critical
slowing.  (Although, as is the case for all cluster algorithms, the
parallel time needed for a single MC step must increases logarithmically
with the system size \cite{MaGr96}). The dynamic exponent $z$ is  defined
by $\tau \sim L^{z}$. We obtained the dynamic exponents $z_m= -.01 \pm
.01$, $z_{\epsilon}=  -.46 \pm .01$ and $z_f= -1.23 \pm .10$ for $m$,
$\epsilon$ and $f$ respectively.

\begin{figure}[h]
\epsfxsize = 3.5in
\begin{center}
\leavevmode
\epsffile{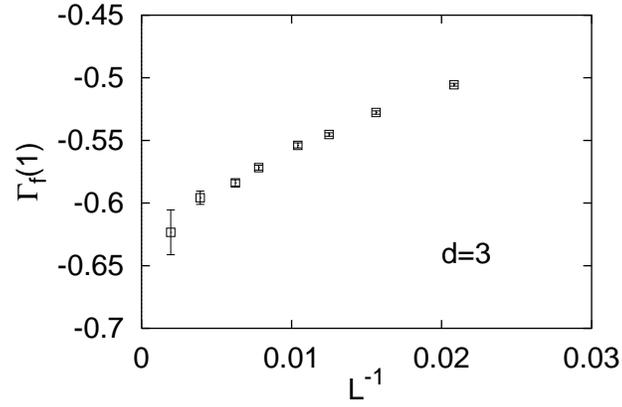}
\caption{Plot of $\Gamma_f(1)$  vs. $L^{-1}$  for the three-dimensional
Ising model.}
\label{fig:fig10}
\end{center}
\end{figure}

Figure \ref{fig:fig10} shows $\Gamma_f(1)$ as a function of $1/L$.  It
appears that the infinite volume limit for $\Gamma_f(1)$ lies between
between $-0.6$ and $-0.65$.  Since $\tau_f \geq 0$ there must be 
corresponding positive values summing to at least $0.1$ to $0.15$ for
$t \geq 2$.   It would be interesting to have more information about the
infinite volume limit of the $f$ autocorrelation function.

The autocorrelation function is related to the standard
error, $\delta A$ in measuring $<\!A\!>$ according to, 
\begin{equation}
\label{eq:errtau}   \delta A= \sigma_A(2/N)^{1/2}( \tau_A -
\frac{1}{N}\sum_{t=1}^{\infty}t\Gamma_A(t))^{1/2}
\end{equation} 
with $\sigma_A^2=<\!A^2\!>-<\!A\!>^2$ and $N$ the
number of Monte Carlo steps. In most Monte Carlo applications the second
term is much smaller than the first however for the IC algorithm this may
not be the case since $\tau_f, \tau_{\epsilon} \rightarrow 0$.  The
autocorrelation function for $f$ is approximately $-1/2$ for $t=1$ and
small for $t>1$ so that the approximate error in measuring the critical
coupling is \begin{equation}
\delta f \approx  \sigma_f(2/N)^{1/2}( \tau_f +
\frac{1}{2N})^{1/2}.
\end{equation}
Suppose that the number of MC steps is large enough that $\tau_f
\gg 1/N$ as is the case here for the smaller system sizes.  Combining the
scaling behavior of $\tau_f$ and of $\sigma_f$ we have, for $N \gg L^u$,
\begin{equation}
\delta f \sim L^{(z_f/2-u)}/N^{1/2} \sim
L^{-1.31}/N^{1/2}. 
\end{equation}
On the other hand, if $\tau_f$ and $1/N$ are comparable as is the case
here for the largest lattices then $\delta f \sim
L^{(z_f-u)} = L^{-1.92}$ and $N \sim L^{-z_f}= L^{1.23}$.  

\section{Conclusions}
\label{sec:con}
We used a parallel implementation of the invaded cluster algorithm
to study two- and three-dimensional Ising models.  We have found that the
parallel algorithm is quite efficient for large system sizes where most
computational work is carried out constructing clusters within processing
nodes and relatively little time is spent in communication.  

We have shown that the invaded cluster algorithm has no critical slowing
down for the three-dimensional Ising model as measured by integrated
autocorrelation times for the magnetization, energy or estimated
critical temperature.  Indeed, the integrated autocorrelation time for
the latter two quantities approaches zero as the system size grows so
that relatively few Monte Carlo steps are needed for high accuracy
estimates of the finite-size critical energy or critical temperature.   

On the other hand, extrapolation to the infinite volume limit for
quantities such as the critical temperature is complicated by the absence
of a well-understood finite-size scaling theory for the invaded cluster
ensemble.  We have examined the hypothesis that the distribution of $f$ is
controlled by a single scale factor and found that this apparently holds
for large systems (greater than $128^3$) but that there are significant
corrections to the predicted asymptotic linear dependence of $<\!f\!>$
on $\sigma_f$.  This is not unexpected since the width of the
$f$ distribution is rather large compared to $|\!<\!f\!>-p(K_c) |$.  Thus
relatively small corrections to scaling for the $f$ distribution
can lead to relatively large deviations from  the proposed linear behavior
of $<\!f\!>$ vs.\ $\sigma_f$.  This means that to confidently
reduce systematic errors in estimating the infinite volume critical
temperature we would either have to go to much larger systems to make
$\sigma_f$ small or we would need a detailed understanding of the
corrections to scaling in the IC ensemble.  Our current estimates of the
critical temperature using the IC method are consistent with the best
conventional estimates but have errors which are about an order of
magnitude larger. 

In contrast to the critical temperature, corrections to scaling in
measuring the critical dimension of the spanning cluster are very small
and we are able to obtain an excellent estimate of the magnetic
exponent.  Our result is $y_h=2.482 \pm .001$($\beta/\nu=0.518\pm .001$).

\acknowledgments

This work was supported in part by NSF Grant DMR-9632898.

\begin{table}
\label{table:data1}
\caption{Numerical data for the three-dimensional Ising model using the
1-span rule.}
\begin{center}
\begin{tabular}{rlllll}
L & $\sigma_f$ & $\langle f\rangle$ & $f_{\rm median}$ & M &
  $\varepsilon$ \\
\hline
16 & .02645(4) & .357367(15) & .356223(33) & 859(2) &
  -1.995544(57) \\
32 & .01599(3) & .357961(11) & .357205(33) & 4830(4) &
  -1.995055(26) \\
48 & .01206(2) & .358106(5) & .357534(15) & 13256(9) &
  -1.995242(23) \\
64 & .00988(2) & .358128(4) & .357621(14) & 27078(25) &
  -1.995289(25) \\
80 & .00855(2) & .358145(3) & .357696(11) & 47019(43) &
  -1.995339(19) \\
96 & .00737(2) & .358144(5) & .357783(17) & 74036(86) &
  -1.995362(33) \\
128 & .00625(3) & .358140(2) & .357851(16) & 151241(208) &
  -1.995377(18) \\
160 & .00540(2) & .358136(3) & .357898(19) & 263220(505) &
  -1.995389(15) \\
256 & .00400(4) & .358130(2) & .357904(21) & 847210(3238) &
  -1.995425(25) \\
512 & .00250(7) & .358105(11) & .358098(36) & 4668684(66195) &
  -1.995340(40) \\
\end{tabular}
\end{center}
\end{table}
\begin{table}\label{table:tau-parallel}
\caption{Integrated autocorrelation times for three-dimensional
Ising models for the IC algorithm. Results
are measured at time step $w=6$.}
\begin{center}
\begin{tabular}{rlll}
\hline
L & $\tau_{\varepsilon}$ &$\tau_{m}$ & $\tau_{f}$\\
\hline
16  & .325(4) & .639(3) & .097(4) \\
32  & .248(5) & .637(5) & .048(6) \\
48  & .201(4) & .628(4) & .028(5) \\
64  & .173(5) & .636(5) & .018(5) \\
80  & .153(5) & .630(5) & .016(6) \\
128 & .107(8) & .618(8) & .008(8) \\
160 & .094(11) & .606(11) & .008(12) \\
\hline
\end{tabular}
\end{center}
\end{table} 
\end{document}